\documentclass[12pt]{article}
\pdfoutput=1
\usepackage[utf8]{inputenc}
\usepackage{amsmath,amssymb,graphicx,epsfig,color,float,cancel,cite}
\usepackage[titletoc,title]{appendix} 
\usepackage[export]{adjustbox}
\usepackage{multicol,tabu,multirow,bm}
\usepackage{booktabs}
\usepackage[colorlinks]{hyperref}
\usepackage{slashed}
\usepackage[normalem]{ulem}
\usepackage{xcolor}
\def\deltaNBBN{\rm \Delta N_{eff}^{BBN}}

\def\nn{\nonumber}
\def\eV{{\rm eV}}
\def\keV{{\, \rm keV}}
\def\MeV{{\, \rm MeV}}
\def\GeV{{\, \rm GeV}}

\def\nn{\nonumber}
\def\bea{\begin{eqnarray}}
\def\eea{\end{eqnarray}}
\def\ba#1\ea{\begin{align}#1\end{align}}

\newcommand{\beq}{\begin{equation}}
\newcommand{\eeq}{\end{equation}}
\def\dis{\displaystyle}
\allowdisplaybreaks

\begin{document}
\thispagestyle{empty}
%
%
\begin{center}
\vspace*{0.5cm}
{\Large\bf Constraints on Axion-Lepton coupling from Big Bang Nucleosynthesis\\}
\bigskip
{\large Diptimoy Ghosh}\,$^{a, 1}$, \, \, 
{\large Divya Sachdeva}\,$^{a, 2}$ \\
\bigskip 
\bigskip
{\small
$^a$ Department of Physics, Indian Institute of Science Education and Research Pune, Pune 411008, India
}
\end{center}
\bigskip 
\vspace*{3cm}
\begin{center} 
{\Large\bf Abstract} 
\end{center}
\vspace*{-0.35in}
\begin{quotation}
\noindent 
{In this article, we study the implications of the coupling between 
Axion-Like-Particles (ALPs) and Leptons to cosmology in particular,
the Big Bang Nucleosynthesis (BBN). 
We show that the BBN, through the constraint on the  effective number of relativistic
neutrino species, provides the most 
stringent bound on the ALP-electron interaction strength for the mass of axion between
20 keV and 1 MeV. For other values of the mass, the BBN bound complements the stellar-evolution and laboratory bounds.}
\end{quotation}
\bigskip
%
%
\vfill
%
%
\bigskip
\hrule
\vspace*{-0.1in}
\hspace*{1mm}
 $^1$ diptimoy.ghosh@iiserpune.ac.in  \hspace*{1.5cm} $^2$ divya.sachdeva@students.iiserpune.ac.in 

\newpage 
\vspace*{-3mm}
\section{Introduction}
The Standard Model of Particle Physics (SM) has been immensely successful; however, there remains a few unsolved 
puzzles, one of them being the smallness of the coefficient $\bar\theta$ which is a sum of two apparently
independent terms in the theory
\begin{equation}
\bar{\theta}\,=\,\theta\,+\,\text{arg det}(Y_u Y_d) \, ,
\end{equation}
where $\theta$ is the coefficient of the classically marginal\footnote{The leading contribution to the beta function of $\bar\theta$ in the 
SM appears at the 7-loop order \cite{Ellis:1978hq,Khriplovich:1993pf}.} and CP odd operator $-(1/32\pi^2) G_{\mu\nu}^a \tilde G^{a \, \mu \nu}$,
and $Y_{u,d}$ are the quark Yukawa matrices. Interestingly, the coefficient $\bar\theta$ contributes to Electric Dipole Moments (EDMs) of the 
leptons \cite{Choi:1990cn,Ghosh:2017uqq} as well as the hadrons, e.g., the neutron \cite{Crewther:1979pi,Baluni:1978rf}.
vacuum and the Yukawa coupling of Higgs boson to quarks, 
The strong experimental constraint on the neutron EDM gives an upper bound $\bar \theta\,\leq\,10^{-10}$~\cite{Afach:2015sja,Crewther:1979pi,Baluni:1978rf}. 
The smallness of this parameter is intriguing, and known as the ``strong CP problem".

The most popular solution of this problem, known as the the Peccei Quinn (PQ) solution
\cite{PhysRevLett.38.1440,PhysRevD.16.1791,PhysRevLett.40.223,PhysRevLett.40.279}, predicts the existence of a light
pseudo Nambu-Goldstone Boson (pNGB), called the axion,  in the particle spectrum. The axion gets its mass from non-perturbative
QCD dynamics, and in this case, the axion mass is completely determined by the axion decay constant denoted by $f_a$, see
\cite{Peccei:2006as,Marsh:2015xka} and the references therein for more details.

Light pNGBs, such as the QCD axion discussed above, appear in many extensions of the SM \cite{Ringwald:2014vqa}.
These particles, being pNGBS, share many properties with the QCD axion. 
However, since their mass, in general, does not arise
from the QCD dynamics, the mass and the decay constant are independent parameters.
These particles are traditionally called the Axion-Like-Particles (ALPs). 

A lot of a theoretical and experimental efforts are underway to search for the ALPs, for example, they are actively being
searched at Light-shining through wall experiments~\cite{Redondo:2010dp,EHRET2010149}, particle colliders~\cite{del_Amo_Sanchez_2011,mimasu2014alps,Lees_2014,Jaeckel_2016,Bauer:2017ris}, 
beam dump experiments~\cite{PhysRevLett.59.755,PhysRevD.38.3375,Essig_2010}, dark matter detection
experiments~\cite{Armengaud_2013,Akerib_2017,Aprile:2020tmw}
as well as various axion helioscopes and haloscopes~\cite{Graham:2015ouw,Bauer:2017ris,DiVecchia:2019ejf}.
There are also many new proposals such as~\cite{Dobrich:2015jyk,Alekhin:2015byh,Abel:2017rtm,Berlin:2019ahk,Wu_2019,Janish:2019dpr,Smorra:2019qfx,Garcon_2019,Dent_2020}
which are expected to cut into unprobed regions of parameter space, and hopefully yield a signal. Along with these,
ALPs are also constrained by a variety of astrophysical and cosmological observations. 
For example, axions can be produced copiously inside Sun, Supernovae, Pulsars, White dwarfs etc., and affect the
cooling time of these objects as well as the photon polarisation. This gives interesting constraints on axion
properties \cite{Raffelt,Jaeckel_2010,Brust:2013ova,Bollig:2020xdr}
(note, however, that the astrophysical bounds often suffer from large uncertainties \cite{Bar:2019ifz,Isern:2020non}).
Similarly, the Cosmic Microwave Background (CMB) power spectrum, and the Big Bang Nucleosynthesis (BBN)
can also be sensitive to the properties of the ALPs, and are used to constrain significant part of the ALP parameter
space \cite{Vysotsky:1978dc,Chang:1993gm,Cadamuro:2010cz,Cadamuro:2011fd,Brust:2013ova,Millea:2015qra,Baumann:2016wac,DEramo:2018vss}.

Most of the experiments mentioned above rely on the axion coupling to photons. The ALP-Lepton coupling has been  
relatively less studied, see however, \cite{Brust:2013ova,Baumann:2016wac,DEramo:2018vss}.
In this work, we perform a detailed analysis of the cosmological implications of ALP-Lepton
interaction, in particular, its impact on the BBN.
The BBN occurred when the temperature of the Universe was between 10 $\keV\, {\rm <}\, \text{T}\, <\, $1 MeV,
and therefore axions with mass $<\, 1$ MeV can affect the physics in this era. The axions can be produced via 
\[ l^{\pm}\gamma\to l^{\pm}a \, \,\,  \text{and} \, \, \,  l^{+}l^-\to \gamma a\]
processes and can lead to non-negligible axion abundance during BBN and contribute to $\Delta \rm N_{\rm eff}$.
In addition, the ALP-electron coupling $c_e/f_a$ can also induce significantly large coupling to 
 photons at one loop, $g_{a\gamma\gamma}\,\sim\,10^{-3}\,c_e/f_a$, for $m_a\,\gtrsim\,m_e$. This extra interaction would allow
 ALPs  to stay in equilibrium for longer providing stronger bounds.

The paper is organised as follows. In Sec. \ref{section:2} we introduce the general effective Lagrangian describing
the ALPs and their couplings to the photon and the SM fermions, and review the present constraints on ALP-photon and 
ALP-lepton couplings. Sec. \ref{section:3} describes how the ALP-Lepton coupling can affect the thermal history of the universe, and 
impact the effective number of neutrinos, $\rm N_{\rm eff}$, measured during BBN. 
In Sec. \ref{section:4}, we numerically solve the Boltzmann equations to find the exact
constraints on ALPs-Lepton couplings as a function of the ALP mass. 
Finally, in Sec. \ref{section:5}, we summarise our main results and conclude.

\section{Axion properties}
\label{section:2}
We use the following Lagrangian density describing the interactions of axions or ALPs to SM particles
\ba
{\cal L} & =  \frac{1}{2}  (\partial_\mu a)^2 + \frac{\alpha_s}{8\pi} \frac{a}{f_a} G_{\mu\nu}^A {\tilde G}^{A \, \mu\nu} + 
\frac{g_{a\gamma\gamma}}{4} a F_{\mu\nu} {\tilde F}^{\mu\nu} + c_\psi \frac{\partial_\mu a}{2 f_a} \bar{\psi} \gamma^\mu \gamma_5 \psi
\label{eq:lagr}
\ea
where $f_a$ is the effective axion decay constant and $\alpha_s$ is the QCD fine structure constant. The axion-photon-photon coupling,
$g_{a\gamma\gamma}$,  is model dependent. 
The coupling $g_{a\gamma\gamma}$ for the QCD axion can be written as
\bea
g_{a\gamma\gamma} &=& \frac{\alpha_{\rm em}}{2 \pi f_a} \left( \frac{E}{N} - 1.92 \right)
\eea
where the quantities $E$ and $N$ are the mixed anomaly coefficients of the PQ symmetry with Electro-Magnetism (EM) and QCD respectively, 
and the number 1.92 is the model-independent contribution from QCD \cite{diCortona:2015ldu}. The mass of the QCD axion is also determined by the
axion decay constant in the following way\cite{diCortona:2015ldu}
\bea
m_a &=&\dis 5.7 \,\mu \eV \,  \left( \frac{10^{12}\GeV}{f_a}\right) \, .
\eea
This gives, 
\bea
g_{a\gamma\gamma} &=&  \left( \frac{E}{N} - 1.92 \right) \left(\frac{m_a}{0.5 \, \rm eV} \right) \, (10^{-10} \GeV^{-1})
\eea

\begin{table}[h!]
\centering
\begin{tabular}{ccc}
\toprule
 $\mathbf{m_a}(\MeV)$  &  $\mathbf{g_{a\gamma\gamma}}$\,(GeV$^{-1}$)  &  Description\\
 \toprule
  $\lesssim 10^{-7}$ & $\lesssim 10^{-7}$ & Light-shining-through-walls experiment~\cite{Redondo:2010dp}\\
  $\lesssim 10^{-5}$ & $\lesssim 10^{-10}$ &Tokyo Axion Helioscope (SUMICO) and \\
  &&the CERN Axion Solar Telescope (CAST)~\cite{Arik:2008mq,Graham:2015ouw}\\  
  $\lesssim 10^{-2}$ & $\lesssim 10^{-10}$ &Horizontal Branch stars~\cite{Raffelt}\\
  $\lesssim 10^{2}$ & $\geq 10^{-2}\,\&\,\lesssim 10^{-9}$ & Supernova Type II SN1987A~\cite{Jaeckel_2010}\\
  $\sim 1 - 10^6$ & $\lesssim 10^{-3}$ & Mono-photon and tri-photon searches at\\
  &&LEP, CDF and LHC~\cite{Jaeckel_2016}\\
  \toprule
 \end{tabular}
\caption{\em Model independent constraints on $g_{a\gamma\gamma}$ for different values of $m_a$. 
Note that the ALP-photon coupling $g_{a\gamma\gamma}$ can also give rise to interesting effects on the CMB properties
which, in turn, can be used to put serious constraints on $g_{a\gamma\gamma}$, see for example, refs.~\cite{Cadamuro:2010cz,Cadamuro:2011fd,Brust:2013ova,Millea:2015qra,Baumann:2016wac,DEramo:2018vss,Depta:2020wmr}.
\label{tab:gayy}}
\end{table}

The current bounds on $g_{a\gamma\gamma}$ and its variation with mass are shown in table~\ref{tab:gayy}. For example,
$g_{a\gamma\gamma} \lesssim 10^{-10} \rm \, \GeV^{-1}$ for {$m_a \lesssim \rm 10\, \keV$} which translates to $m_a \lesssim 0.5 \,  \rm \eV$ 
(and consequently, $f_a \gtrsim 10^7 \GeV$) for QCD axion with $(E/N -1.92) \sim 1$. 
The decay width of the ALP decaying to photons can be written as, 
\bea
\Gamma_{a \to \gamma \gamma} = \left( \frac{g_{a\gamma\gamma}}{4} \right)^2 \, \frac{m_a^3}{4 \pi} &=&  \left( \frac{g_{a\gamma\gamma}}{10^{-10} \GeV^{-1}} \right)^2
\left( \frac{m_a}{350 \, \rm eV} \right)^3 \left( \frac{1}{3.3 \times 10^{17} \rm sec.} \right) \label{eqn:Adecaygaga}\\
  &&\hspace{-15mm} =
\left(\frac{E}{N} - 1.92 \right)^2 \, \left( \frac{m_a}{25.5 \, \rm eV} \right)^5 \left( \frac{1}{3.3 \times 10^{17} \rm sec.} \right)\, [\text{for QCD axion}]
\eea
Thus, for the QCD axion, one necessary condition for it to have a lifetime more than the age of the universe is that the axion mass has to be less
than about 25 eV (for (E/N - 1.92) $\sim$ 1). For a general ALP, its lifetime can be larger than the age of the universe even when the mass is much larger
than 25 eV, and depends on $g_{a\gamma\gamma}$ (and also couplings to matter) which is not determined by $m_a$.

\begin{figure}[H]
\begin{center}
\begin{tabular}{cc}
 \hspace{-3mm}\includegraphics[width=6.9cm, height=5.2cm]{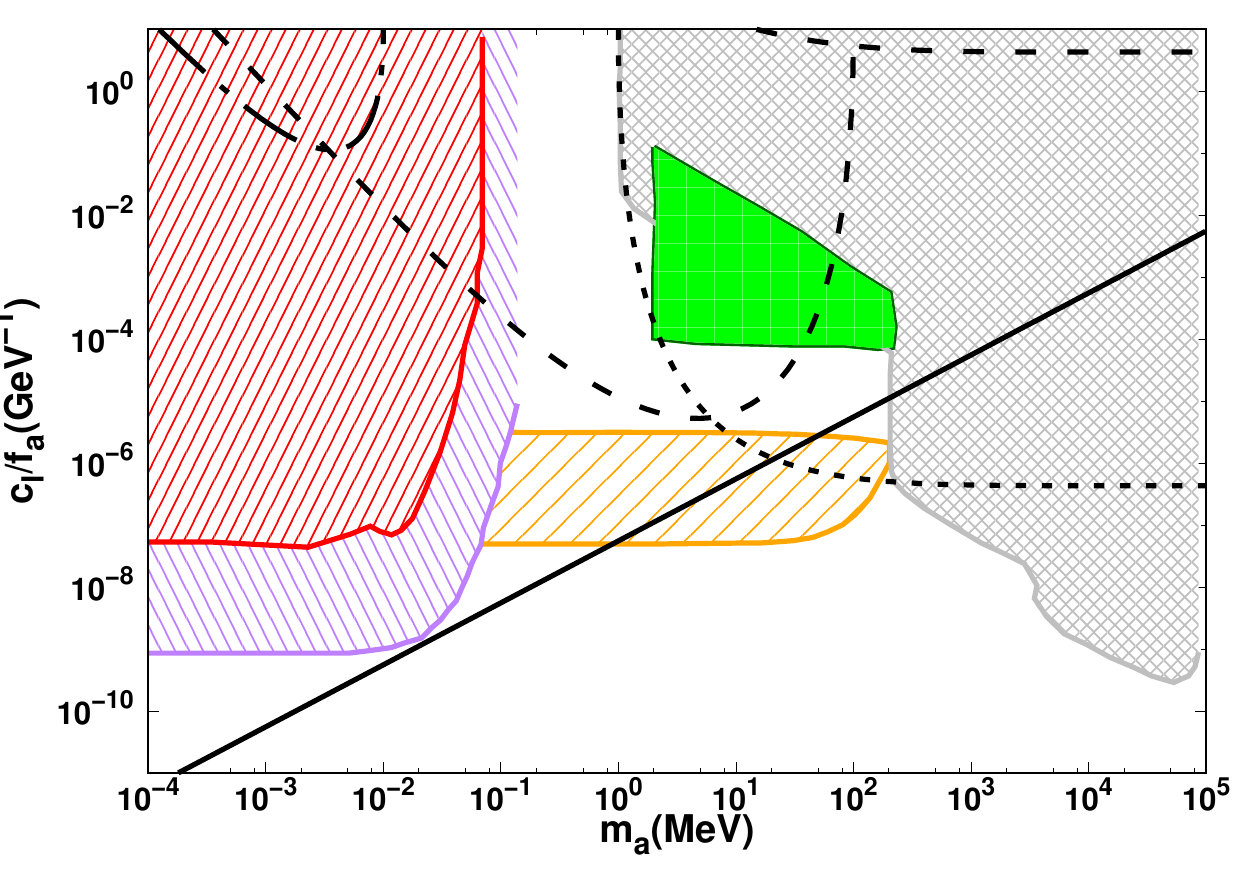} &
  \includegraphics[width=9.1cm, height=5.2cm]{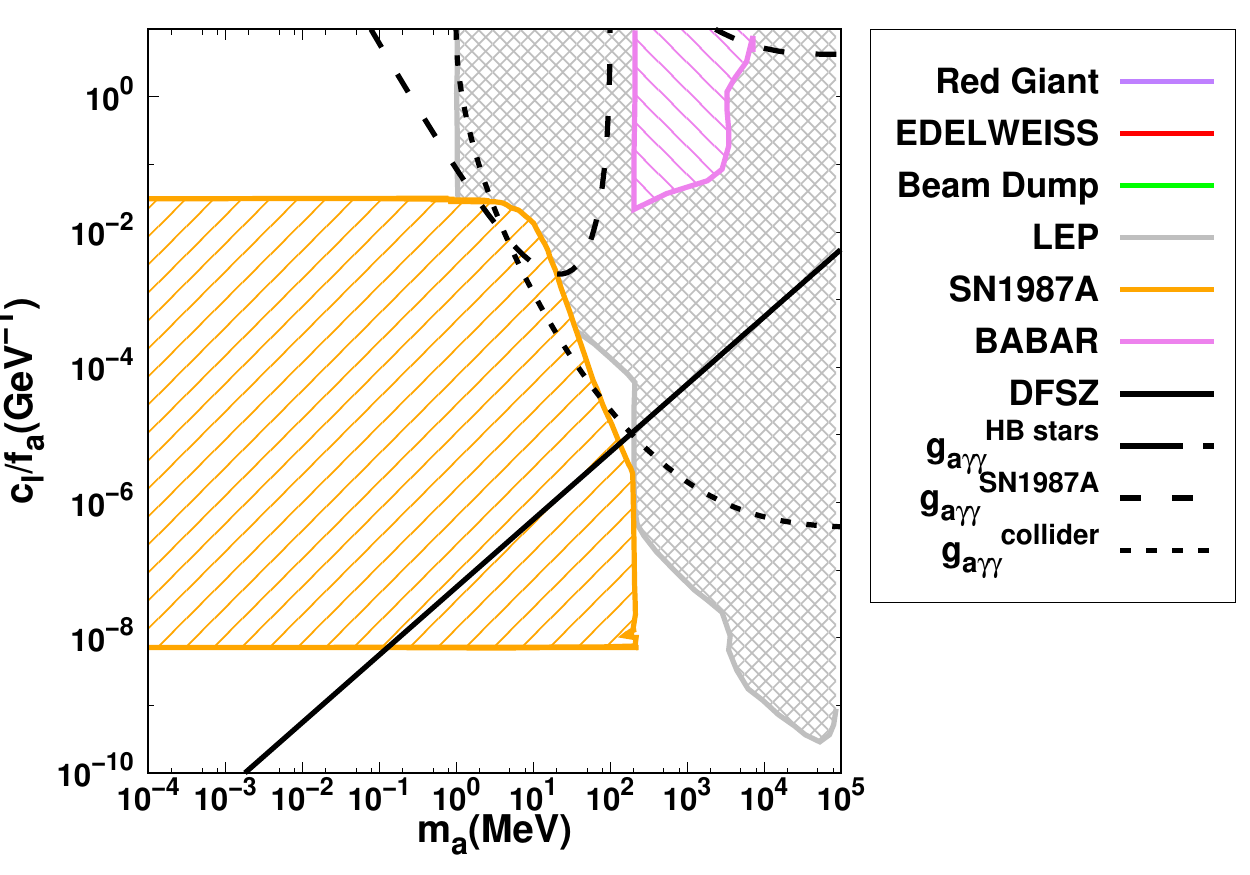}
  \end{tabular}
 \caption{ \em Existing model independent constraints on $c_e/f_a$ {\em (left)} and $c_\mu/f_a$ {\em (right)} for different
 values for $m_a$ from observations of Red Giants~\cite{Raffelt,Calibbi:2020jvd}, SN1987A~\cite{Brust:2013ova,Bollig:2020xdr,Croon:2020lrf,Calibbi:2020jvd}, dark photon searches at 
 BabaR~\cite{Lees_2014}, Beam dump experiment at SLAC~\cite{Essig_2010}. Note that collider bounds from LEP~\cite{Bauer:2017ris} assumes 
 $g_{a\gamma\gamma}\sim 10^{-3}\,\GeV^{-1}$, so they are 
 not model independent. Bounds from EDELWEISS~\cite{Armengaud_2013} and LUX~\cite{Akerib_2017} are one order of magnitude stronger than shown here,
 if axion are assumed as cold dark matter. Similarly, BBN studies in \cite{Brust:2013ova} constrains 
 $(c_e/f_a, c_\mu/f_a, c_\tau/f_a) \sim (10^{-6},10^{-7},10^{-6})$ assuming relativistic axions 
 during CMB decoupling i.e, with $m_a\,<\,10^{-1}\eV$. 
 The CAST experiment \cite{Barth_2013} constraints the product $g_{a\gamma\gamma} c_e/f_a < 1.6 \times10^{-19} \GeV^{-2}$
 at 95\% CL for $m_a\lesssim 10^{-2}$eV. 
 The solid black diagonal line corresponds to the DFSZ QCD axion model. }
\label{fig:constr1}
\end{center}
\end{figure}

Similar to $g_{a\gamma\gamma}$, the axion-fermion-fermion couplings, $c_\psi$, is also model dependent. 
In this work, we consider couplings only to the charged leptons, $e$, $\mu$ and $\tau$ (couplings to the neutrinos are extremely suppressed due
to smallness of the neutrino masses). 
In models such as the
Dine-Fischler-Srednicki-Zhitnitsky (DFSZ) model \cite{Zhitnitsky:1980tq,DINE1981199}, axions have tree-level interactions with leptons and $c_l$ can be
as high as 1/3. In the Kim-Shifman-Vainshtein-Zakharov (KSVZ) model \cite{PhysRevLett.43.103,SHIFMAN1980493}, on the other hand, 
such direct interaction with leptons is absent.
Even in models where there is no tree level coupling to leptons, it can be generated by the axion-photon-photon coupling at one loop.
As the axion-photon-photon coupling already has one power of $\alpha_{\rm em}$, $c_l$, generated in this way, is $\mathcal{O} (\alpha_{\rm em}^2)$
suppressed. However, for QCD axions, large logarithms either of the { size ${\rm Ln}(\Lambda_{\rm QCD}/m_e) \approx 5$ or ${\rm Ln}(f_a/m_e) \approx 30$ may} kill
part of this suppression. In Fig.~\ref{fig:constr1}, we compile all the existing bounds on $c_l/f_a$ for different values of $m_a$. 

\begin{figure}[!ht!]
\begin{center}
\begin{tabular}{cc}
\hspace{-1.5cm}\includegraphics[scale=0.55]{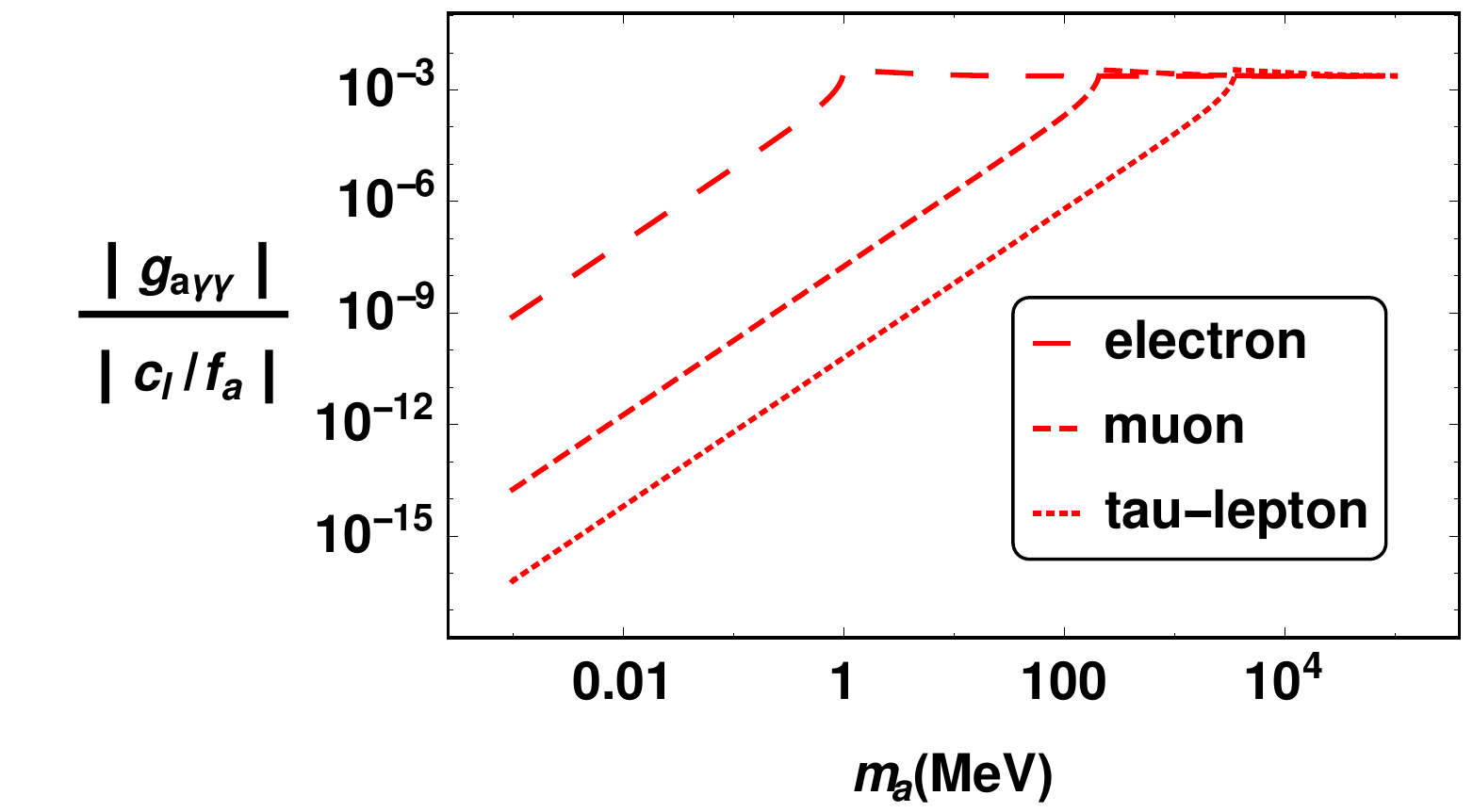} &
\includegraphics[scale=0.55]{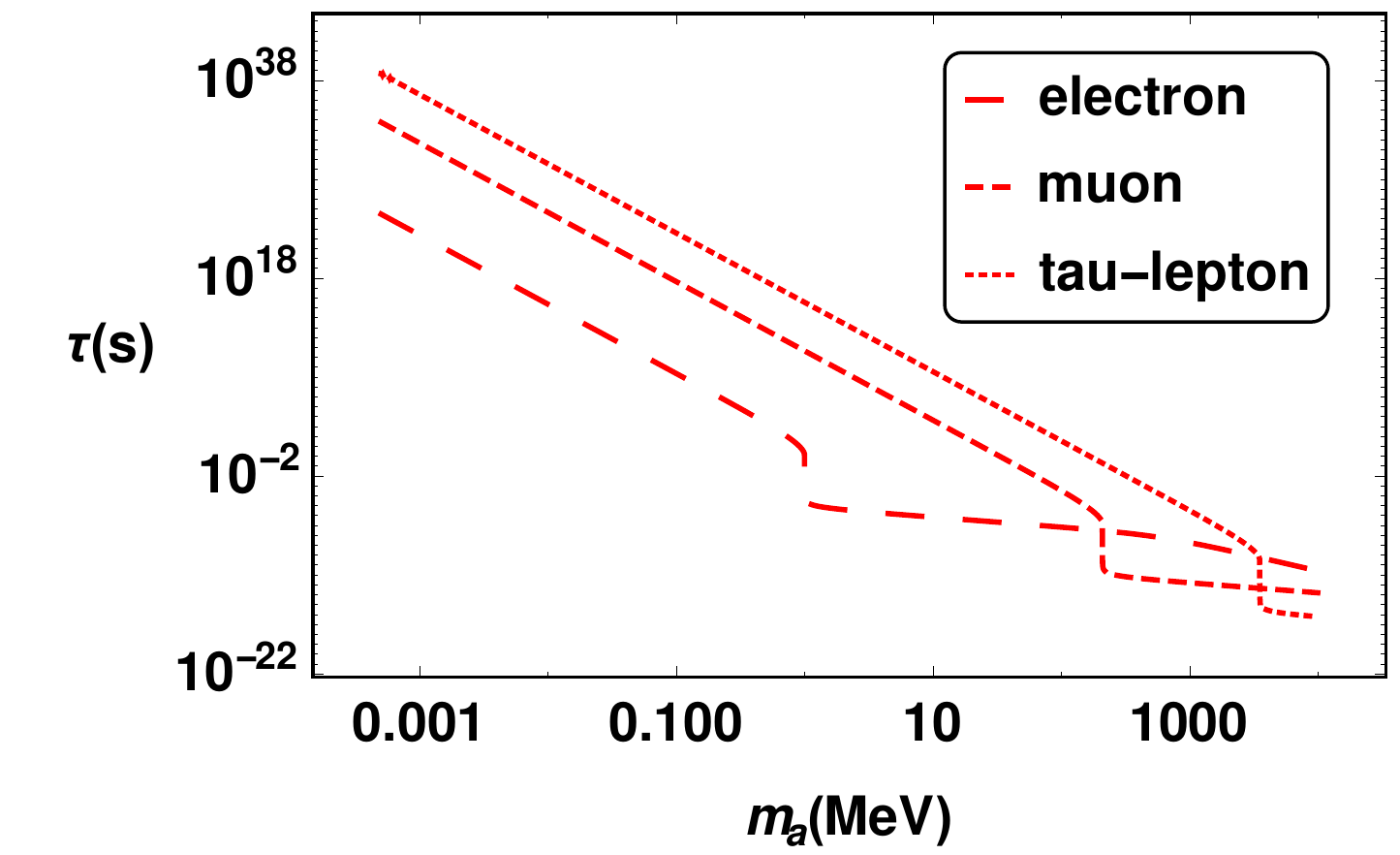} 
\end{tabular}
\caption{\em (left) The variation of ratio $\dis\left|\frac{g_{a\gamma\gamma}}{c_l/f_a}\right|$ with $m_a$. (right) Lifetime of axion decaying to photon as 
well as leptons as a function of mass for {$c_l/f_a\,=\,10^{-4}\,\GeV^{-1}$.}
\label{fig:gagaga_loop}}
\end{center}
\end{figure}

As we wish to extract constraints on the couplings $c_l/f_a$ in a model independent way, we assume that the direct
coupling to photons is zero (the bounds on $c_l/f_a$ will only get stronger for non-zero $g_{a\gamma\gamma}$, thus our bounds are conservative). 
Note, however, that the ALP-Lepton coupling does generate the axion photon coupling, $g_{a\gamma\gamma}$, at one loop:
\bea
g_{a\gamma\gamma}^\text{loop}\,&=&\,\dis   \frac{\alpha_{\rm em}}{4 \pi} \, \frac{c_l}{f_a} \, 4 \,B_1(x_l)\label{eq:gayy}\\
\text{where,} \qquad 
B_1(x_l)\,&=&\,1-x_l\,f\left(x_l\right)^2\,; \qquad x_l\,\equiv\,\frac{4m_l^2}{m_a^2}\nn\\
f(x)\,&=&\, \left\{ 
   \begin{array}{ll} 
    \dis\arcsin\frac{1}{\sqrt{x}} \,; &~ x \ge 1 \,, \nn \\
    \dis \frac{\pi}{2} + \frac{i}{2} \ln\frac{1+\sqrt{1-x}}{1-\sqrt{1-x}} \,; &~ x < 1 \,. 
   \end{array} \right.
\eea
The fermion loop function $B_1 (x) \,\sim\,-\frac{m_a^2}{12\,m_l^2}$ for $m_l \gg m_a$ and $|B_1(x)|\,\sim\,1$ for $m_a \gg m_l$.
We plot $g_{a\gamma\gamma}^\text{loop}$ as a function 
of the ALP mass for different leptons in Fig.~\ref{fig:gagaga_loop}({\em left}).

It is evident that the value of $g_{a\gamma\gamma}$ is non-negligible  especially for $m_a\gg m_l$ with $g_{a\gamma\gamma}\stackrel{m_a\, \gg\, m_l}{\sim}\dis \frac{\alpha_{\rm em}}{\pi} \, \frac{c_l}{f_a}$.
To emphasize its significance, we plot lifetime of ALPs as function of their mass keeping $c_l/f_a\,=\,10^{-4}\GeV^{-1}$ in Fig.~\ref{fig:gagaga_loop}({\em right}).
To calculate the lifetime, we use eqn.~\ref{eqn:Adecaygaga} and the following expression for decay rate of
the ALP to leptons, 
\bea
\Gamma_{a \to l^+ l^-} = \frac{m_l^2}{8\pi}\left( \frac{c_l}{f_a} \right)^2 \, m_a \,\sqrt{1-\frac{4m_l^2}{m_a^2}}. 
\eea

\section{Axions production in the early universe}
\label{section:3}
Considering only the axion-lepton-lepton coupling, $c_l$, the axions can be generated in the early universe by the following processes
\bea l^{\pm}(p_1)\,\gamma(p_2)&\to l^{\pm}(p_3)\,a(p_4) \label{eqn:process1}\\
l^{-}(p_1)\,l^+(p_2)&\to \gamma(p_3)\, a(p_4).
\label{eqn:process2}
\eea
\begin{figure}[!ht!]
\begin{center}
\begin{tabular}{cc}
 \includegraphics{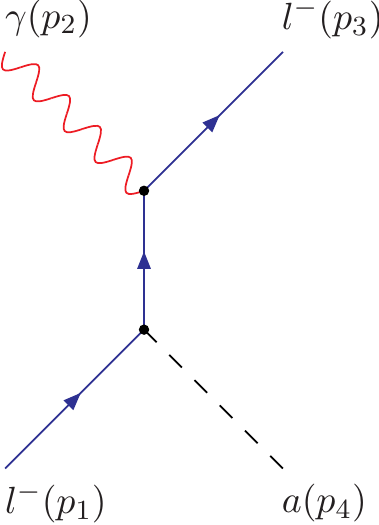} &
  \includegraphics{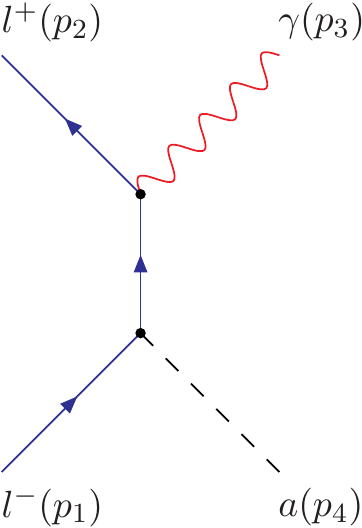} 
  \end{tabular}
 \caption{\em Feynman diagrams of $l^{\pm}(p_1)\,\gamma(p_2)\to l^{\pm}(p_3)\,a(p_4)$ and $l^{+}(p_1)\,l^-(p_2)\to \gamma(p_3)\, a(p_4)$}
  \label{feyn-diag}
  
\end{center}
\end{figure}
The corresponding Feynman diagrams are shown in Fig.~\ref{feyn-diag}.
Based on the dimensional analysis, one expects $\langle \sigma v\rangle$ to scale as $\dis \frac{c_l^2m_l^2}{f_a^2\,T^2}$ in the relativistic limit (for $T \,\gg\,m_{l,a}$). 
This implies that the ratio of interaction rate to Hubble expansion takes the form 
$\dis \frac{\Gamma}{H}\sim \frac{n_{l,\gamma}\langle \sigma v\rangle\,M_\text{pl}}{T^2}\propto \frac{1}{T}$. As Universe cools to 
temperatures below the lepton mass, $\langle \sigma v\rangle$ goes as $\dis \frac{c_l^2m_l^2}{f_a^2}\frac{1}{g(m_a^2,m_l^2)}$ where
$g(m_a^2,m_l^2)$ is 
 a function having mass dimension 2. Consequently, $\Gamma/H$ would go as $\dfrac{n_{\gamma}\langle \sigma v\rangle\,M_\text{pl}}{T^2}\propto T$. However, the detailed analysis discussed below suggests that the interaction rate in the 
non-relativistic limit drop more quickly than expected from the naive power counting. For that, we write below the 
Boltzmann equation describing the evolution of the axion number density
\bea
 \frac{dn_a}{dt} \,+\, 3\,H\,n_a\,&=&\, \left(2\Gamma^{l^\pm\gamma\to a\,l^\pm} + \Gamma^{l^-l^+\to a\gamma}\right)(n_a-n^\text{eq}_a)\nonumber\\
\qquad\frac{dY_a}{dx}\,&=&\,-\frac{x\,s}{H(m)}
\left(2\langle\sigma v\rangle^{l^\pm\gamma\to a\,l^\pm}Y_l^\text{eq}\frac{Y_l^\text{eq}}{Y_a^\text{eq}}\,+\,\langle\sigma v\rangle^{l^-l^+\to a\gamma}\,Y_\gamma^\text{eq}\frac{Y_l^\text{eq}}{Y_a^\text{eq}}\right)\,
(Y_a\,-\,Y_a^\text{eq})\nonumber \\
\label{eqn:boltzmann}
\eea
where, $Y_a = \displaystyle \frac{n_a}{s}$, and $x =\displaystyle\frac{m_l}{T} $. The matrix element for process in eqn.\ref{eqn:process1} gets two 
contributions associated to s- and u-channel exchange. The corresponding expressions are given as: 
\bea
i\mathcal{M}^{s}\,&=&\,\dis \frac{i\,e\,c_l}{2f_a}\,\bar{u}(p_3)\,\gamma^\mu\gamma^5\frac{i(\slashed{p_1}-\slashed{p_2}+m_l)}{s-m_l^2}\gamma^\nu\, u(p_1)\,p_{4\mu}\,\epsilon_\nu(p_2),\nonumber \\
i\mathcal{M}^{u}\,&=&\,\dis \frac{i\,e\,c_l}{2f_a}\,\bar{u}(p_3)\,\gamma^\nu\frac{i(\slashed{p_1}-\slashed{p_4}+m_l)}{u-m_l^2}\gamma^\mu\gamma^5\, u(p_1)\,p_{4\mu}\,\epsilon_\nu(p_2),\nonumber
\eea

Similar expression for the process in eqn.\ref{eqn:process2} can be derived using crossing symmetry property. We see that the transformation of the Mandelstam 
variables from $(s,t,u) \,\to\, (t,s,u)$ gives us the matrix element for pair-annihilation process.

Next, we calculate the cross-section in the limit $m_a\to 0$, useful in relativistc limit ($T \gg m_{l,a}$) 
\bea
\sigma^{l^\pm\gamma\to l^\pm\,a}\,&=&\,\frac{\alpha c_l^2 m_l^2}{8 f_a^2 s^2}\frac{ \left(-m_l^4+4 m_l^2 s+ s^2 \log \left(\frac{s^2}{m_l^4}\right)-3 s^2\right)}{ \left(s-m_l^2\right)}\nonumber\\
&\stackrel{s\, \gg\, m_l^2}{=}&\,\dis\frac{\alpha c_l^2 m_l^2}{8 f_a^2 s} \left(\log (s^2/m_l^4) - 3\right)\label{eqn:xsection1}\\
\sigma^{l^+l^-\to \gamma\,a}\,&=&\frac{\alpha c_l^2 m_l^2}{4f_a^2} \frac{\tanh ^{-1}\left(\sqrt{1-\frac{4 m_l^2}{s}}\right)}{(s-4 m_l^2)}\nonumber\\
&\stackrel{s\, \gg\, m_l^2}{=}&\,\dis\frac{\alpha c_l^2 m_l^2}{4f_a^2\,s}\tanh ^{-1}\left(\sqrt{1-\frac{4 m_l^2}{s}}\right)\label{eqn:xsection2}
\eea
It should be noted that the logarithmic dependence leads to divergent cross-section in the $m_l \,\to \, 0$ limit corroborating the t-channel singularity.

\begin{figure}[!ht!]
\begin{center}
\begin{tabular}{cc}
\hspace{-5mm}\includegraphics[scale=0.5]{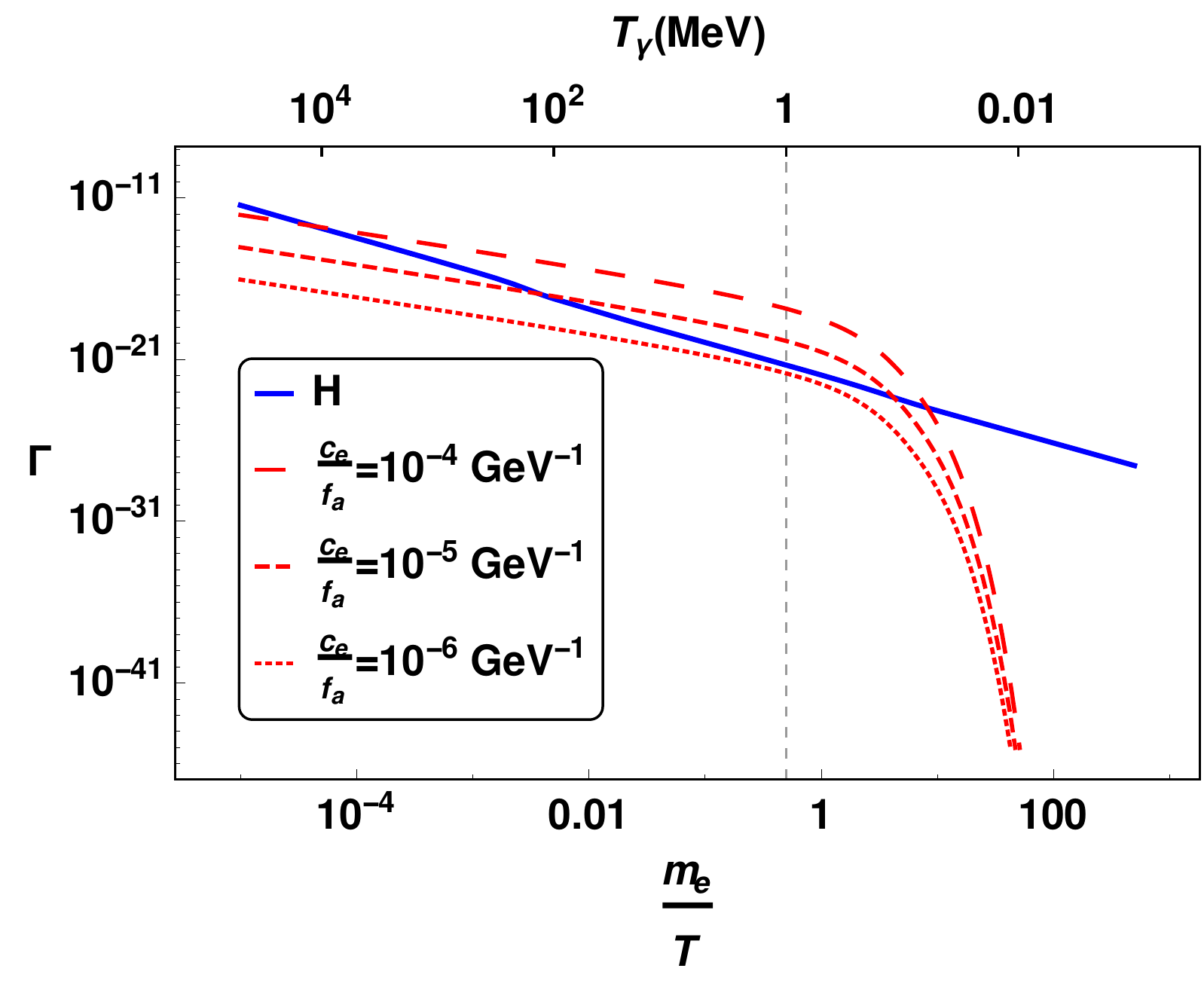} &
 \includegraphics[scale=0.5]{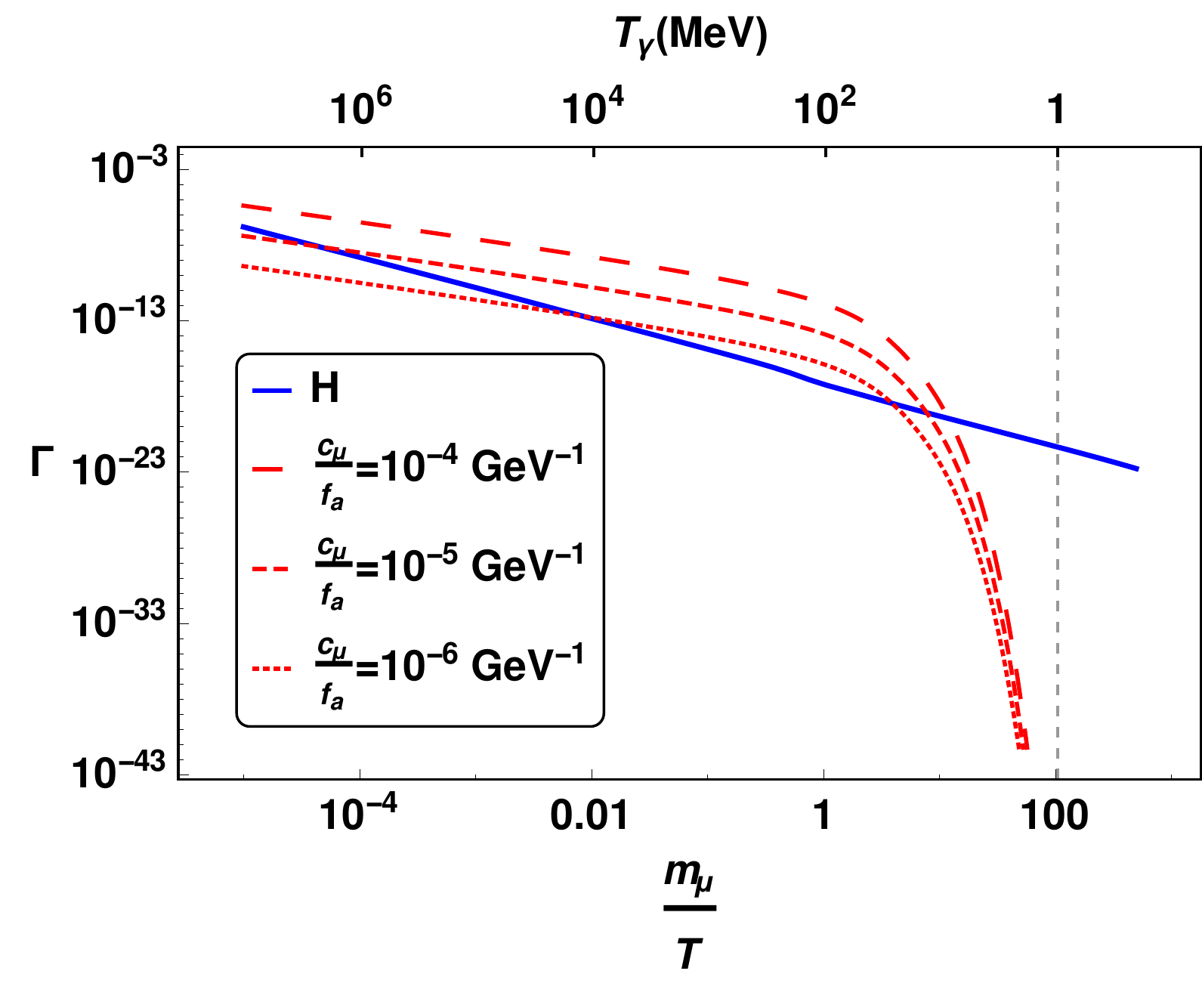} \\
\multicolumn{2}{c}{\includegraphics[scale=0.5]{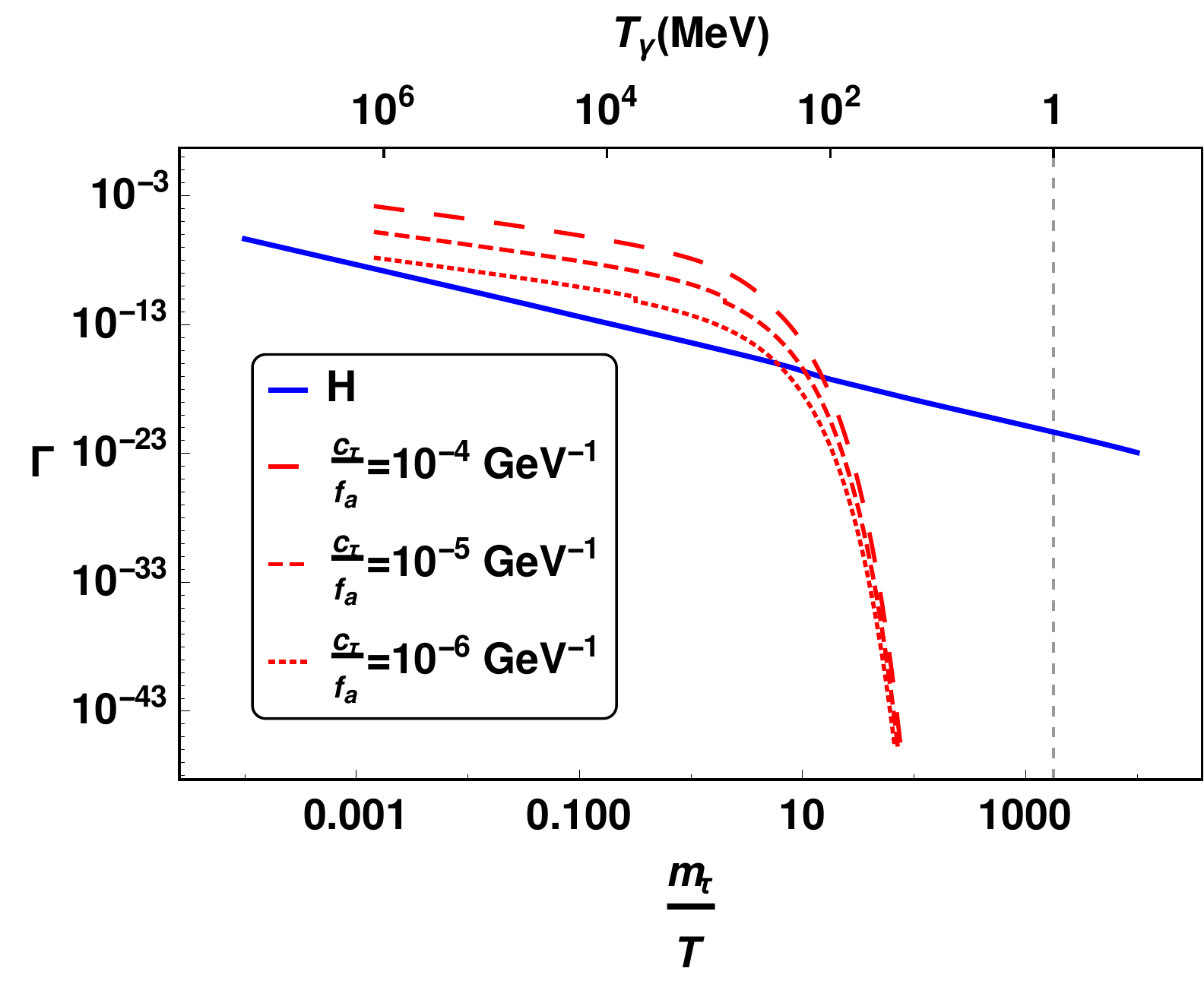}}
 \end{tabular}
 \caption{\em Interaction rate and Hubble rate as function of temperature for $m_a = 10\,{\rm keV}$ and for various coupling strength.}
 \label{fig:GammaToH}
\end{center}
\end{figure}

From eqn.\ref{eqn:boltzmann}, \ref{eqn:xsection1} and \ref{eqn:xsection2}, the reaction rate per axion particle in the limit $T\gg m_{a,l}$ can be written as 
\bea
\Gamma\,=\,\frac{n_{l/\gamma}^{\rm eq}n_{l}^{\rm eq}}{n_a^{\rm eq}}\langle \sigma v\rangle\, &\stackrel{\text{eqn.} \ref{eqn:process1},\ref{eqn:process2}}{\sim} &\,  T^3  \,\left( \frac{\alpha_{\rm em}}{16}\,\frac{c_l^2\,m_l^2}{f_a^2}
\, \frac{1}{T^2}\text{Log}\left[\frac{4T^2}{m_l^2}\right] \right) \,=\, \frac{\alpha_{\rm em}\, T}{16}\,\frac{c_l^2\,m_l^2}{f_a^2}
\,\text{Log}\left[\frac{4T^2}{m_l^2}\right].\nonumber
\eea
Since, the Hubble parameter, $H \sim {T^2}/{\text{M}_\text{pl}}$,
\bea
\frac{\Gamma}{H} &\propto& \frac{1}{T}\text{Log}\left[\frac{4T^2}{m_l^2}\right]
\eea
Thus, as the universe cools, $\Gamma/H$ keeps growing allowing the possibility to bring the axions in thermal equilibrium with the plasma. 
However, when the universe cools further so that $T \ll m_{l}$, one expects
\bea
\Gamma&\stackrel{\text{eqn.} \ref{eqn:process1}}{=}&\frac{n_{\gamma}^{\rm eq}n_{l}^{\rm eq}}{n_a^{\rm eq}}\langle \sigma v\rangle\, \sim\,
T^3 \, \frac{n_{l}^{\rm eq}}{n_a^{\rm eq}}  \, 
\left[\frac{ \alpha_{\rm em}}{2m_l^2} \, \frac{c_l^2m_l^2}{f_a^2} \, \left(a\,\frac{T}{m_l}+b\,\frac{T^2}{m_l^2}\right) \right]\\
&\stackrel{\text{eqn.} \ref{eqn:process2}}{=}&\frac{n_{l}^{\rm eq}n_{l}^{\rm eq}}{n_a^{\rm eq}}\langle \sigma v\rangle\, \sim \, e^{-m_l/T}\,(m_lT)^{3/2}\, 
\frac{n_{l}^{\rm eq}}{n_a^{\rm eq}} \, 
\left[\frac{ \alpha_{\rm em}}{2m_l^2} \, \frac{c_l^2m_l^2}{f_a^2} \, \left(\tilde{a} +\tilde{b}\,\frac{T}{m_l}\right) \right].
\eea
where $a,\tilde{a},b$ and $\tilde{b}$ are functions of the axion and the lepton masses and are found numerically to be $\mathcal{O}(1)$ numbers. 
It should be noted that the interaction in equation 
\ref{eqn:process1} is dominant at lower temperature as the other interaction has an extra Boltzmann suppression factor corresponding to the extra 
lepton in the initial state. Thus, the ratio $\Gamma/H$ scales (other than the Boltzmann suppression) as $T^n$ 
with $n\,=\,2$ in the non-relativistic regime i.e it falls quicker than expected within dimensional analysis. We also conclude that this ratio attains
its maximum value around $T\,\sim\, m_l$. This feature can be seen in figure \ref{fig:GammaToH}. Note that this feature arises only for axion-lepton-lepton couplings because it is effectively a marginal coupling. It can be seen from the figure that the axion
produced in this way stays in thermal equilibrium (i.e., $\Gamma > H$) for some duration. Depending on the coupling and mass, it goes out of the 
equilibrium at lower temperatures. For ALPs interacting with only the muons or the tau-leptons, they go out of equilibrium before
the neutrino decoupling; however, they can still contributes significantly to the total energy budget of the Universe. In fact, as we show in the next section, the 
 energy density stored in the ALPs could be significant enough to affect the BBN. 
The non-negligible yield and hence, the energy density of axions can increase the Hubble parameter during the BBN.
A larger Hubble parameter during the BBN modifies the neutron-to-proton ratio, which in turn changes the abundance of
Helium-4 and Deuterium. This effect can be captured by defining a quantity called $\deltaNBBN$ in the following way. 
The total energy density during the BBN can be written as
\bea
\rho&=& \frac{\pi^2}{30}\{g_\gamma + \frac{7}{8}\left( 3g_\nu + 3g_{\bar{\nu}}+ g_{e^+}+g_{e^-}\right)\}T_\gamma^4 + \rho_a \\
&=&\frac{\pi^2}{30} \frac{43}{4} T_\gamma^4  + \rho_a \\
&=& \frac{43}{8} \rho_\gamma  + \rho_a
\eea
The quantity $\deltaNBBN$ can now be defined by 
\bea
\rho &=& \frac{\pi^2}{30} \left( \frac{43}{4} + \frac{7}{8} \, 2 \, \deltaNBBN \right)T_\gamma^4   \\
\implies \deltaNBBN &=& \frac{8}{7} \, \frac{\rho_a}{\rho_\gamma} \label{deltaNeff-formula}
\eea
The latest measurement and analysis of Helium and Deuterium abundance constrain 
this parameter to $N_{\rm eff}^{\rm BBN}\,=\, 2.878\pm{0.278}$ at 68.3\% CL~\cite{Fields:2019pfx}. Using
$\rm N_\text{eff}= 3.046$ for the SM \cite{Mangano:2005cc,Akita:2020szl}, this gives $\deltaNBBN < 0.39$ at $2\sigma$.
Since, some of the earlier works~\cite{Mangano:2011ar,Cooke:2013cba,Cyburt:2015mya,Pitrou:2018cgg,Berlin:2019pbq}
had a slightly weaker upper bound, in order to be very conservative we take $\deltaNBBN\,<\,0.5$ in our work.

\begin{figure}[!ht!]
\centering
\begin{tabular}{cc}
\hspace{-5mm}\includegraphics[width=0.5\linewidth]{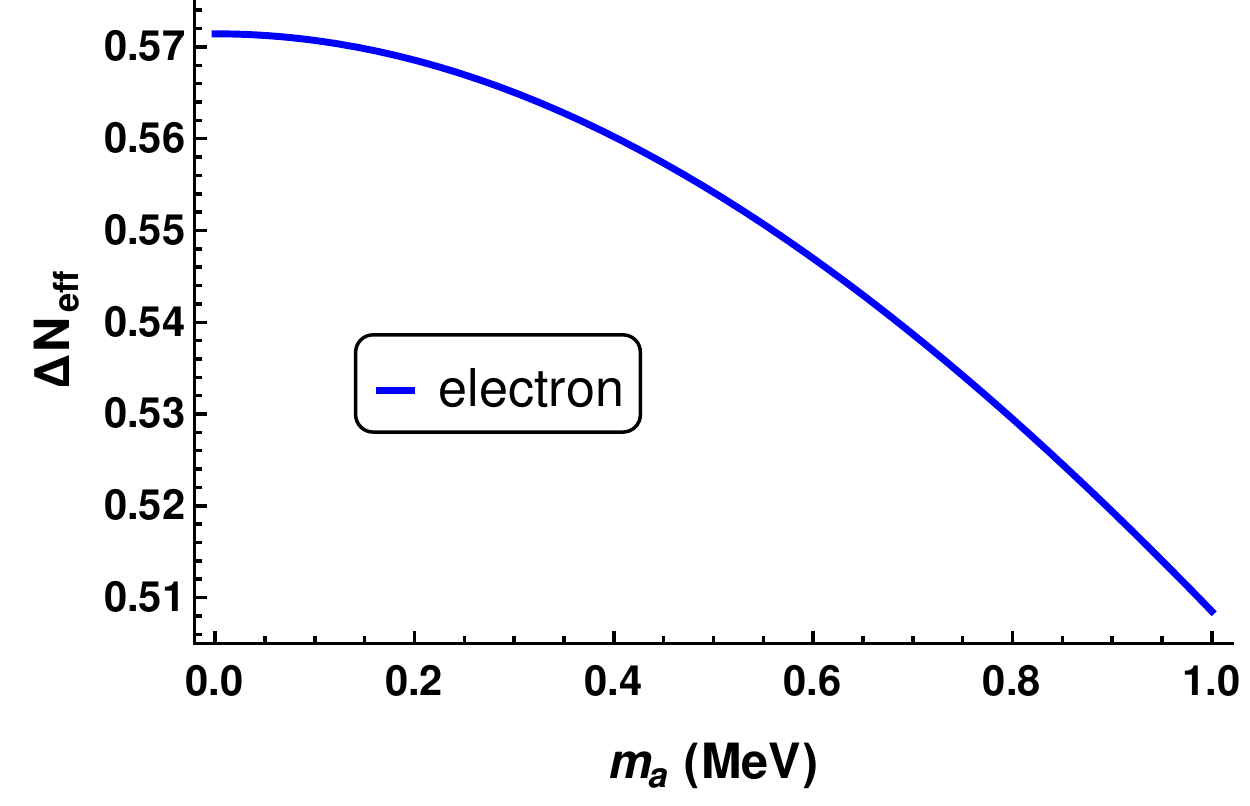} & 
\includegraphics[width=0.5\linewidth]{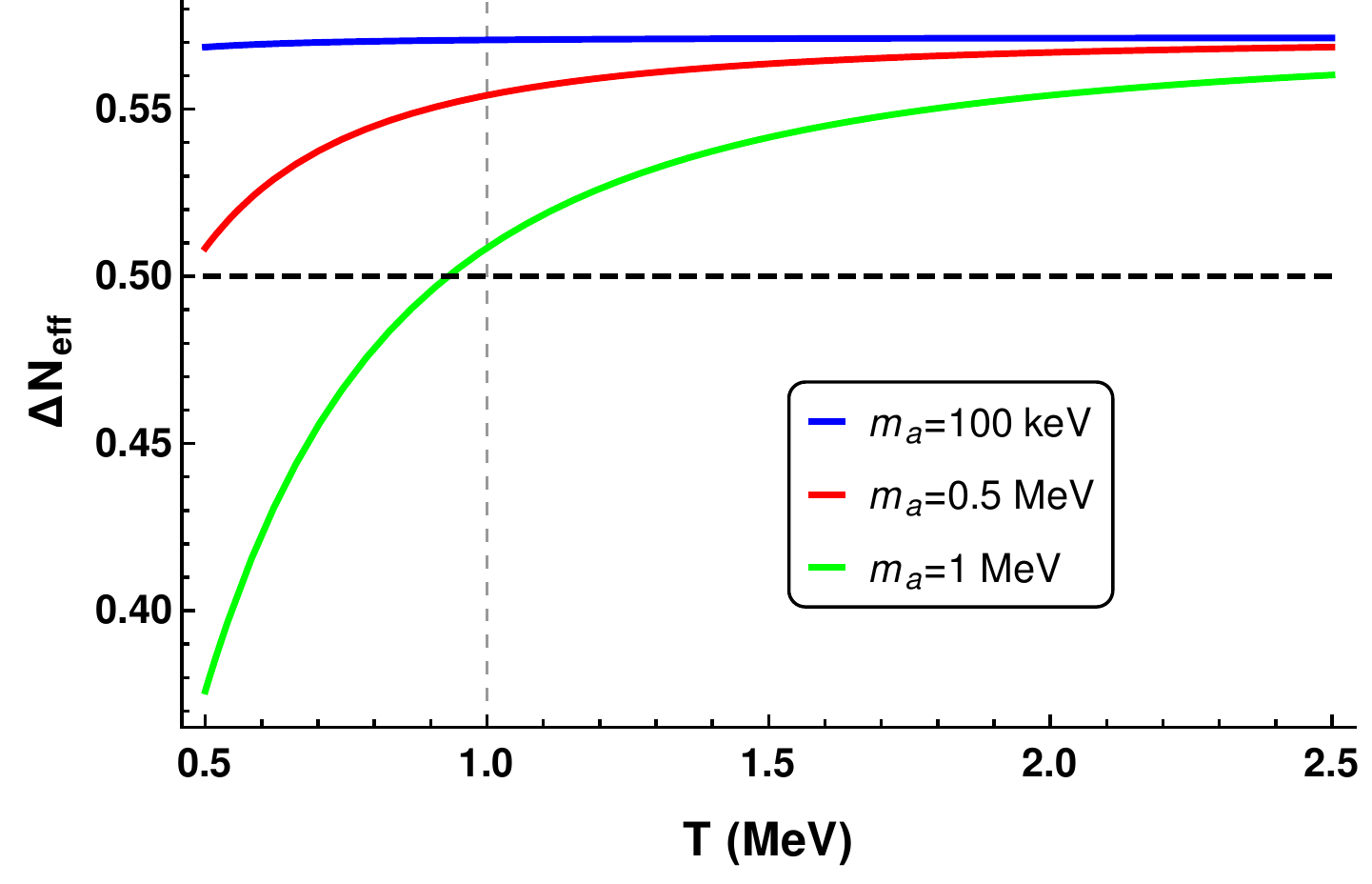} 
\end{tabular}
\caption{$\Delta N_{\rm eff}$ as function of $m_a$ {\em (left)} and $T$ {\em (right)} assuming that the ALPs are in thermal equilibrium.
In the left panel, only the ALP-electron coupling has been considered.}
\label{fig:Neff-vs-ma}
\end{figure}

The maximum value of $\deltaNBBN$ in the context of ALPs is obtained if the ALPs stay in thermal equilibrium during the
BBN and are relativistic (i.e., $m_a \ll 1 \MeV$). In this case, $\deltaNBBN = 8/7 \times 1/2 = 0.57$ since the axion,
being a pseudo-scalar, contributes only one extra degree of freedom. For larger axion mass, $\deltaNBBN$ reduces
from 0.57 owing to the the Boltzmann suppression, as can be seen from Fig.~\ref{fig:Neff-vs-ma} (left). In the right panel of
Fig.~\ref{fig:Neff-vs-ma}, we also show $\deltaNBBN$ as a function of the temperature for three different values of $m_a$.
Note that Fig.~\ref{fig:Neff-vs-ma} shows $\deltaNBBN$ for particle at thermal equilibrium. In order to fully take into
account various processes when ALPs are not in thermal equilibrium, we need to solve a complete set of Boltzmann 
equations for the distribution functions of ALPs as well as neutrinos. This will be discussed in the next section.

%

We would like to stress here that the quantity $\Delta \rm N_{\rm eff}$ is also measured from the CMB observations.
However, in general, $\Delta \rm N_{\rm eff}^{\rm CMB}$ is not the same as $\Delta \rm N_{\rm eff}^{\rm BBN}$; depending on the
details, $\Delta \rm N_{eff}^{CMB}$ can be more or less than $\Delta \rm N_{eff}^{BBN}$ \cite{Fischler:2010xz}.
In order for the axions to contribute to $\Delta \rm N_{eff}^{CMB}$, they have to be relativistic at the time of
CMB decoupling. This means that for the axion mass $m_a \gtrsim \eV$, they would not contribute significantly to
$\Delta \rm N_{eff}^{CMB}$. This is why we have not used the stronger constraint on $\Delta \rm N_{eff}$
obtained from the CMB\cite{Aghanim:2018eyx,Fields:2019pfx}. The decay of the ALPs after BBN and close to the CMB decoupling can, however, lead to
observable effects on radiation density and also on other observables like CMB spectral distortions which can
in principle give rise to interesting bounds \cite{Hou_2013,Acharya:2019uba}.

\section{Constraint on $c_l/f_a$}
\label{section:4}
In the homogeneous and isotropic Universe with the FLRW metric, the relevant
distribution functions fulfill the following Boltzmann equations:
\[\frac{\partial f_i(|\vec{p}|,t)}{\partial t}-H |\vec{p}| \frac{\partial f_i(|\vec{p}|,t)}{\partial |\vec{p}|}=C[f_i(|\vec{p}|,t)\]
To solve these first-order partial differential equations, we adopt characteristics curves method~\cite{riley_hobson_bence_2002} and introduce 
two dimensionless parameters $z=m_l \,R(t)$ and $Q = |\vec{p}|\,R(t)$, where $R(t)$ is scale factor in the FLRW metric. So, the
Boltzmann equations is recast into the following form: 
\beq H z\frac{\partial f_i(Q,z)}{\partial z}=C[f_i(Q,z)] \, ,
\label{eqn:dist}\eeq
where the subscript $i$ stands for $e^{\pm},\, \nu \,\text{and} \, a $. While
photons and electrons initially follow the distributions in thermal equilibrium with zero chemical potentials, 
the initial abundance of axion is assumed to be negligible. The collision term for the ALPs can be written as:
\[C[f_\text{ALPs}]=\frac{1}{2E_a}\int dp_ldp'_ldp_\gamma(2\pi)^4\delta^4(\sum p)\left[f_\gamma f_l(1+f_a)(1-f'_l)-f'_lf_a(1-f_l)(1+f_\gamma)\right]|\mathcal{M}|^2\]
where $dp_i=g_id^3p_i/[(2\pi)^32E_i]$ with $g_i$ being the internal degrees of freedom. The collision terms for the SM species are taken from references~\cite{Hannestad:1995rs,Dolgov:1997mb,Esposito:2000hi}.
We use numerical techniques discussed in~\cite{Hannestad:1995rs,Huang:2017egl} to solve multi-dimensional collision
integration. 

\begin{figure}[h!]
\begin{center}
 \includegraphics[width=7.7cm, height=7cm]{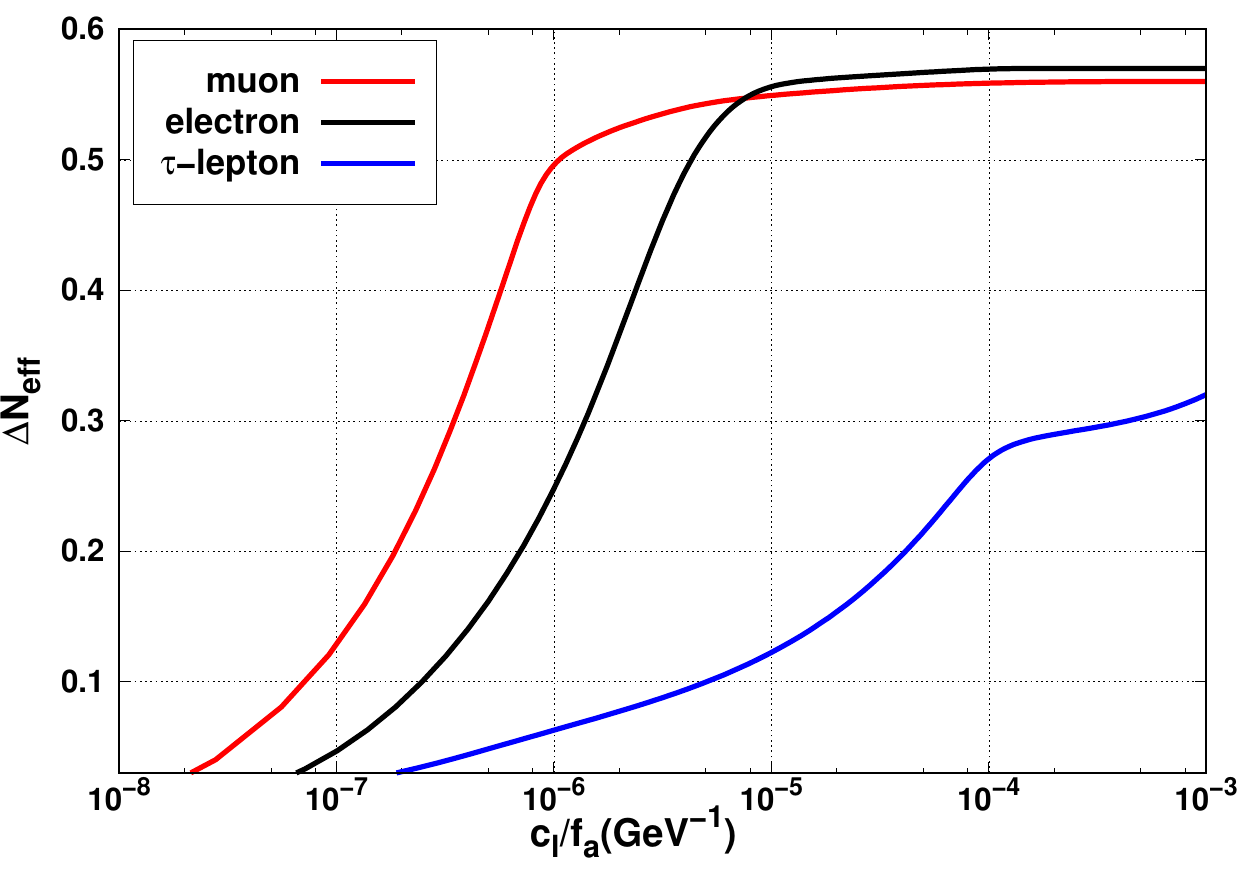} 
 \includegraphics[width=7.7cm, height=7cm]{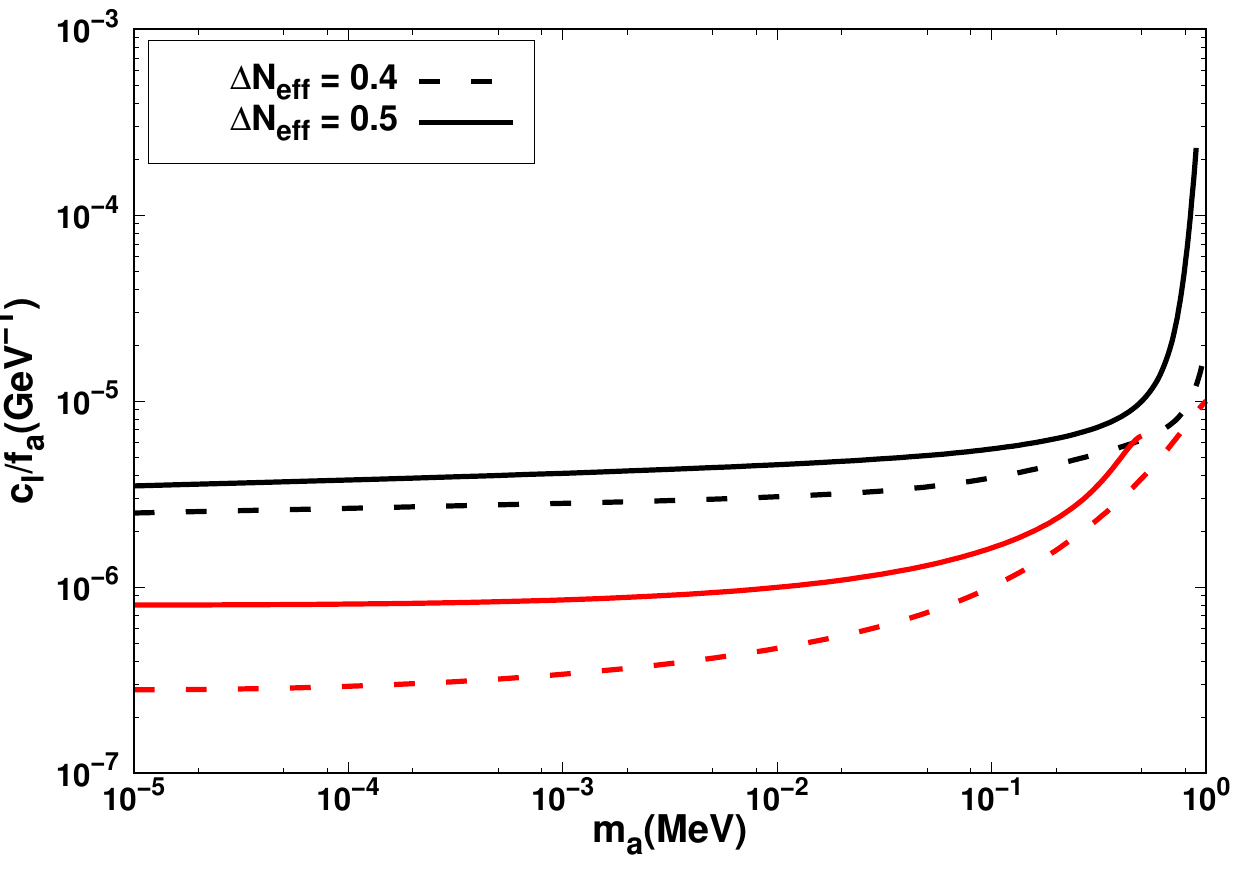} 
 \caption{\em (left) $\Delta N_{eff}$ as a function of $c_l/f_a$ for $m_a \,<\, 10\,$ keV. 
 (right) Contour satisfying $\Delta N_\text{eff.}=0.5$ on $m_a$ vs $c_l/f$ plane.}
  \label{fig:Neff1}
\end{center}
\end{figure}

Along with the Boltzmann equation, the Friedmann equations, $H^2 \,=\,\displaystyle \frac{8\pi G\,\rho^{\rm tot.}}{3}$ and
$\displaystyle\frac{d\rho}{dt}\,=\,\displaystyle -3{\rm H}(\rho + {\rm P})$ in terms of $z$ and $Q$
are also relevant to calculate the evolution of photon temperature:
\ba
z\frac{d\rho}{dz}\,&=\,-3(\rho + {\rm P})\\
z\frac{dT_\gamma}{dz}\frac{d\rho_{\gamma}}{dT_\gamma}&=\,-3(\rho + {\rm P})-z\frac{d\rho_e^\pm}{dz}-z\frac{d\rho_\nu}{dz}-z\frac{d\rho_a}{dz}
\ea
After solving the Boltzmann equation, energy density during BBN and hence $\Delta N_{\text{eff}}$ are calculated using Eq.~\eqref{deltaNeff-formula}
at the time of neutrino decoupling. 
In Fig.~\ref{fig:Neff1}{\em (left)} we show the variation of $\Delta \rm N_\text{eff}$ with respect to the coupling $c_l/f_a$ 
for axions with mass $m_a\,<\,10\,\keV$ (we have chosen 10 keV just as a benchmark. Also, for this mass range, the axions are relativistic and
$\Delta \rm N_\text{eff}$ is independent of the mass). Fig.~\ref{fig:Neff1}{\em (right)}, on the other hand, shows the contours for  $\Delta \rm N_\text{eff.}=0.5 \, \text{and} \, 0.4$ on the
 $m_a$ -- $c_l/f_a$ plane. Note that we have restricted our analysis to $m_a\,\leq\,1\,{\rm MeV}$. However, in principle, the
 analysis can be extended to heavier axion masses, but the constraints are expected to be much weaker.

Let us now try to understand qualitatively the results in Fig.~\ref{fig:Neff1}. First, note that the ALP-Lepton coupling is proportional to the ALP mass.
Thus, for a given $c_l/f_a$, axions will have stronger coupling to heavier leptons. This means that, in order to get the same axion yield, a relatively
smaller value of $c_\mu/f_a$ would be required  than $c_e/f_a$. Consequently, the constraint on $c_\mu/f_a$ will be stronger than
that on $c_e/f_a$ (for this argument, we have assumed the mass of the axion to be much less than the electron and the muon mass),
which can be seen from the figure. 

For the heavier leptons, however, another effect comes into play: lower number density of leptons for heavier leptons (due to
Boltzmann suppression) leads to lower rate for the process $l^-/\gamma\,+\,a\to l^-l^+$ reducing the yield. In order to maintain
significant axion yield, this requires larger $c_l/f_a$. In this way, the competition between the number density of
 leptons and lepton mass dependent coupling strength plays a crucial role in obtaining the
 constraints given in Fig.~\ref{fig:Neff1}. 
 In the right panel of Fig.~\ref{fig:Neff1}, the increase of required coupling strength with the axion mass in order to maintain
a fixed $\Delta N_\text{eff}$ is also easily understood in terms of the Boltzmann and phase space suppression of the
process $l^-l^+ \to  l^-/\gamma\,+\,a$.

As discussed above, the coupling of the ALPs with tau-lepton cannot be constrained by BBN as the contribution to $\Delta\,N_\text{eff}$
is very small (Fig.~\ref{fig:Neff1}{\em (left)}). In this case, the axions go out of equilibrium much before BBN
(as can be seen from the lower panel of Fig.~\ref{fig:GammaToH}) and thus, their yield remains small.
{This result is consistent with the observation that, for a particle that decouples from the primordial thermal plasma before the QCD transition,
one always gets $\Delta\,\rm N_\text{eff}\,\lesssim \,0.3$~\cite{Brust:2013ova}.

Finally, in Fig.~\ref{fig:constrFin} we compare our bounds with the existing constraints discussed in section.~\ref{section:2}.
It can be seen that, for the ALP-electron coupling, our bound is the strongest for axion mass between 20 keV and 1 MeV.
We should also mention in passing that, indirect bound on the ALPs-lepton coupling can be obtained using Eq.~\eqref{eq:gayy} and the bounds
on ALPs-photon coupling given in table \ref{tab:gayy} (shown as dashed and dotted black lines in Fig.~\ref{fig:constrFin}).
However, unlike the BBN bound discussed in this work, this indirect bound is model dependent and relies on the exact value of the
direct $g_{a\gamma\gamma}$ coupling.

\begin{figure}[H]
\begin{center}
\begin{tabular}{c}
 \includegraphics[width=16cm, height=7cm]{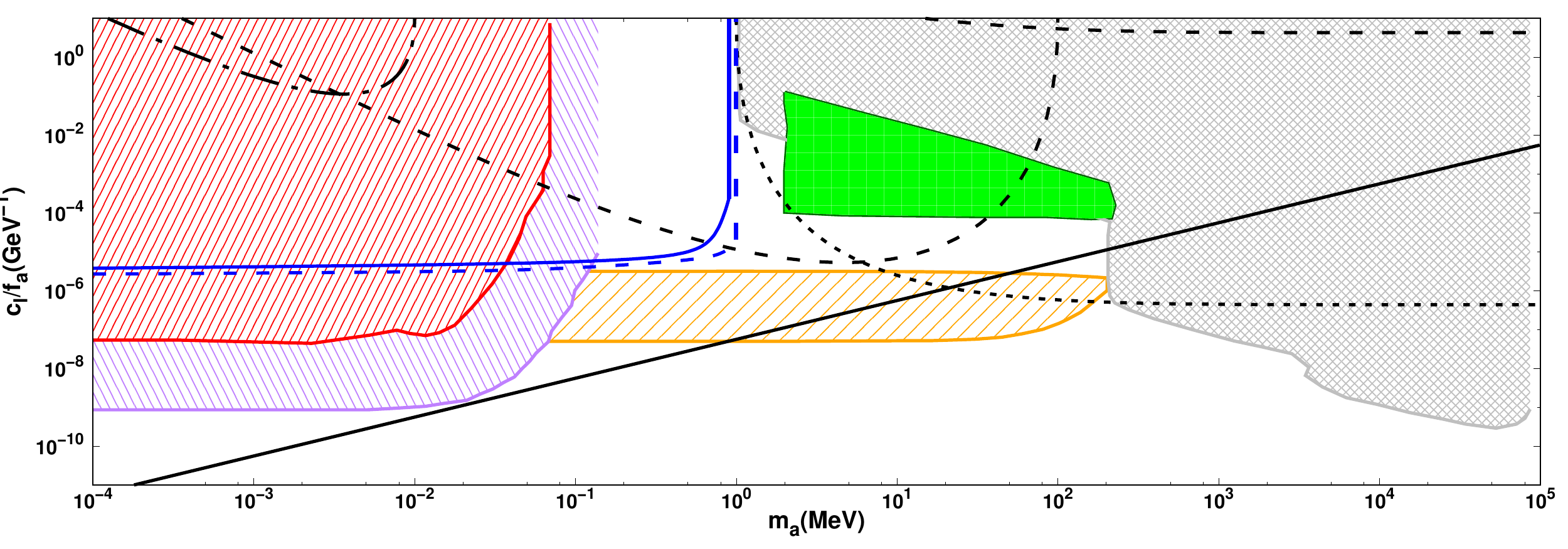} \\
 \includegraphics[width=16cm, height=7cm]{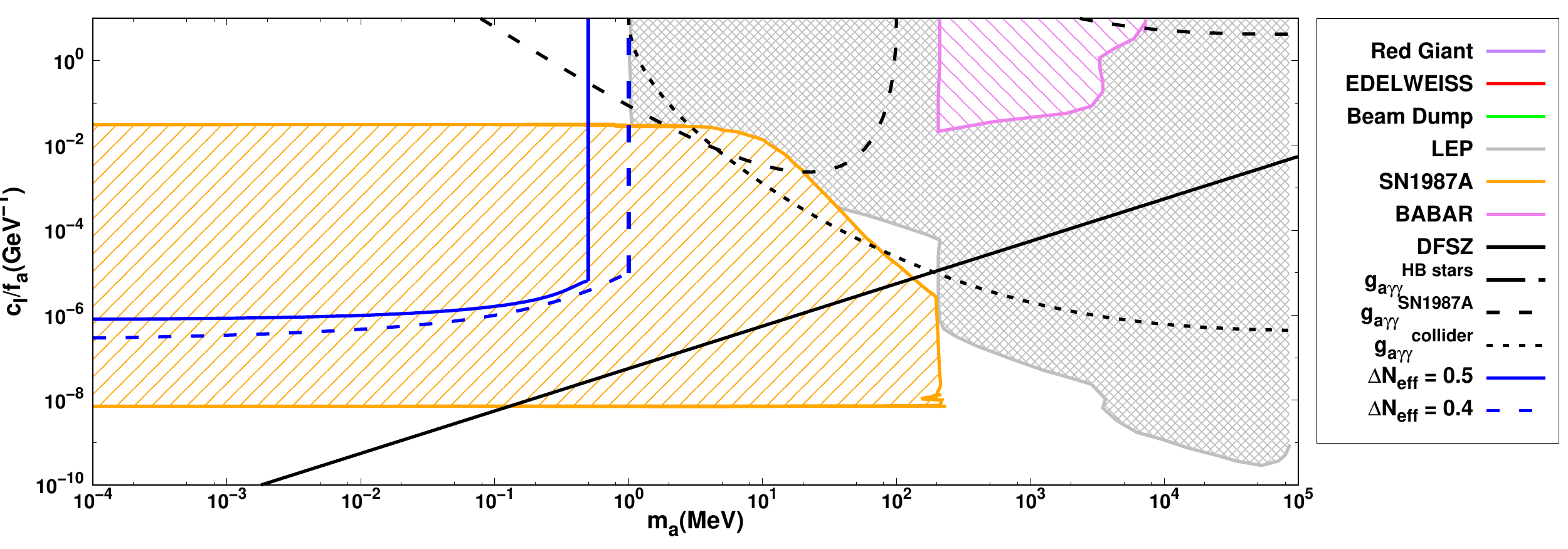} 
  \end{tabular}
 \caption{\em Constraints on $c_e/f_a$ {\em (upper panel)} and $c_\mu/f_a$ {\em (lower panel)} obtained in the present work along with the existing
 model independent constraints. We also show constraints derived using bounds on
 axion-photon couplings discussed in Tab.~\ref{tab:gayy}.
 }
\label{fig:constrFin}
\end{center}
\end{figure}

\section{Summary and Conclusion}
\label{section:5}
In this article, we have performed a detailed study of the constraints on ALP-lepton coupling, 
$c_l/f_a$, from BBN. In the presence of non-zero $c_l/f_a$, the ALPs can be produced in the early universe
by the processes $l^{\pm}\gamma\to l^{\pm}a \, \,\,  \text{and} \, \, \,  l^{+}l^-\to \gamma \, a$ (Fig.~\ref{feyn-diag}).
When the temperature of the thermal plasma $T\gg m_{a,l}$, the quantity $\Gamma/H$ keeps increasing
as the universe cools allowing the possibility for the axions to come into equilibrium with the rest of the plasma. 
However, as the universe cools further and the temperature goes below the mass of the lepton,
$\Gamma/H$ starts decreasing with the decreasing temperature. As a result, the ALPs, depending on their coupling strengths,
first come into equilibrium, then stay in equilibrium for a while, and eventually goes out of the equilibrium
(see Fig.~\ref{fig:GammaToH}). If the axions are relativistic and also in equilibrium during the BBN, their contribution
to $\Delta \rm N_{eff}^{BBN}$ turns out to be as high as 0.57 which is clearly ruled out by observations.
With the increasing mass of the axion, their contribution to $\Delta \rm N_{eff}^{BBN}$ keeps decreasing due to the Boltzmann
suppression (shown in Fig.~\ref{fig:Neff-vs-ma}). However, they can still contribute to $\Delta \rm N_{eff}^{BBN}$
significantly which, in turn, provides interesting constraints on the ALP-Lepton couplings.

When ALPs are not in equilibrium with the thermal plasma, the full Boltzmann equations have to be solved in order to
understand their effect on BBN quantitatively. We did this exercise in section \ref{section:4}. Based on our analysis, in Fig.~\ref{fig:Neff1}, 
we showed contours for $\Delta \rm N_{eff}^{BBN} = 0.5$ on the $c_{e, \mu}/f_a$ -- $m_a$ plane. We also discussed qualitatively
why the constraints on $c_\mu/f_a$ is stronger than the bound  on $c_e/f_a$, and why the bound on $c_\tau/f_a$ is extremely poor. 
Finally, in Fig.~\ref{fig:constrFin} we compare our bound (using a very conservative observational upper limit $\Delta \rm N_{eff}^{BBN} \leq 0.5$)
with the other existing bounds and show that our bound is the most
stringent one for the ALP-electron interaction strength for the mass of axion between
20 keV and 1 MeV.

\section*{Acknowledgement}
The authors acknowledge support through the Ramanujan Fellowship of the Department of Science and Technology, Government of India. 
\begin{appendices}
\end{appendices}
\providecommand{\href}[2]{#2}\begingroup\raggedright\endgroup

\end{document}